# Magnetic and thermal transport properties of SrFe$_{12}$O$_{19}$ permanent magnets with anisotropic grain structure

A. D. Volodchenkov[1,2,3], S. Ramirez[4,5], R. Samnakay[4,5], R. Salgado[4], Y. Kodera[1,2], A. A. Balandin[3,4,5] and J. E. Garay[1,2,3*]

[1]Advanced Material Processing and Synthesis (AMPS) Laboratory, Department of Mechanical and Aerospace Engineering, University of California – San Diego, San Diego, California 92093
[2]Department of Mechanical Engineering, University of California – Riverside, Riverside, California 92521
[3]Spins and Heat in Nanoscale Electronic Systems (SHINES) Center, University of California – Riverside, Riverside, California 92521
[4]Phonon Optimized Engineered Materials (POEM) Center, Department of Electrical and Computer Engineering, University of California – Riverside, Riverside, California 92521
[5]Materials Science and Engineering Program, University of California – Riverside, Riverside, California 92521
*Corresponding author

## Abstract

Nanostructured permanent magnets are gaining increasing interest and importance for applications such as generators and motors. Thermal management is a key concern since performance of permanent magnets decreases with temperature. We investigated the magnetic and thermal transport properties of rare-earth free nanostructured SrFe$_{12}$O$_{19}$ magnets produced by the current activated pressure assisted densification. The synthesized magnets have aligned grains such that their magnetic easy axis is perpendicular to their largest surface area to maximize their magnetic performance. The SrFe$_{12}$O$_{19}$ magnets have fine grain sizes in the cross-plane direction and substantially larger grain sizes in the in-plane direction. It was found that this microstructure results in approximately a factor of two higher thermal conductivity in the in-plane direction, providing an opportunity for effective cooling. The phonons are the dominant heat carriers in this type of permanent magnets near room temperature. Temperature and direction dependent thermal conductivity measurements indicate that both Umklapp and grain boundary scattering are important in the in-plane direction, where the characteristic grain size is relatively large, while grain boundary scattering dominates the cross-plane thermal transport. The investigated nano/microstructural design strategy should translate well to other material systems and thus have important implications for thermal management of nanostructured permanent magnets.

1. Introduction

Permanent magnets (PMs) are used in a wide variety of applications ranging from power sources to switches and actuators [1, 2]. Today, high-performance sintered magnets [3] are playing an increasingly important role in electric motors and generators [4] since PM based designs are usually more efficient than traditional electro-magnet based designs because they do not require external power to generate the requisite magnetic field. In addition, PM based energy generating systems are typically more compact and require less maintenance, and therefore offer lower operational costs than traditional motors and generators.

Despite these promising attributes, there are fundamental problems with the generators and motors that use PMs. The first is the large amounts of PM material – sometimes kg quantities – which are needed for intended applications. Currently, the best PMs rely on rare earth elements whose supply chain in uncertain [1]. One of the most promising strategies for reducing the rare earth content is using the exchange spring concept [5-7] that requires nanostructuring of the PM material. This dovetails into the second problem, which is the need for thermal management, since generator and motors are often intended for use at high temperatures. PM performance is well known to decrease with temperature, ultimately disappearing at the Curie temperature, $T_c$. A solution to retaining the magnetic performance could be an active cooling of the magnets. However, nanostructuring reduces cooling efficiency since the abundance of interfaces in nanostructured materials significantly lowers the thermal conductivity, $k$, of materials [8, 9].

Here, we propose a strategy based on aligned, high aspect ratio grains that allows for nanostructured grains in one direction, while still providing a high $k$ path for efficient cooing. **Figure 1a** illustrates the benefits of an anisotropic, high aspect ratio microstructure. Magnets with grains aligned in such a way that their easy magnetic axis is along the cross-plane, will produce high magnetic field, allowing for stacking of magnets and maximizing surface area of magnetization. Larger grain sizes in the in-plane direction ensure fewer grain boundaries and longer mean free path for heat carriers, maximizing $k$. In order to test these ideas, we produced fine grain sized permanent magnets using current activated pressure assisted densification (CAPAD) [10] that has been shown to be a viable approach for producing exchange coupled magnets [11]. We chose to use strontium ferrite, $SrFe_{12}O_{19}$, henceforth referred to as SFO as a model PM system because of its abundance, low toxicity and high usage in consumer devices [2, 12].

In the rest of the paper we describe the magnetic and thermal properties of nanostructured SFO. In addition to facilitating the proposed cooling strategy, the produced high aspect ratio, anisotropic samples provide an efficient platform for evaluating the effects of grain sizes and porosity on thermal conductivity of this class of materials. Specifically, they allowed us to explore thermal transport in materials with two different length scales but with the same global porosity and the same chemical and phase composition. We used Laser Flash and Hot Disk methods, proven transient techniques for measuring thermal diffusivity and thermal conductivity of bulk materials.

## 2. Experimental Procedures

### 2. 1 Material synthesis and processing

In order to investigate the effect of porosity and grain size on the magnetic and thermal properties of PM materials, nano-scale ceramic magnetic powder was densified into bulk samples using CAPAD technique. Commercial SFO powder with the reported 500 nm particle size (Nanostructured & Amorphous Materials Inc.) was loaded into a graphite die and plunger (19 mm inner diameter) assembly. Mechanical load and current was simultaneously applied in order to achieve heating rates up to ~300°C/min and mechanical pressure up to 140 MPa. Holding temperatures were between 800°C and 1000°C. The mass density of the samples, $\rho$, was measured geometrically and relative density, $\rho_{rel}$, was found using the theoretical density of SFO, 5.1 g/cm$^3$.

### 2.2. Microstructural characterization

Fracture surfaces of densified samples were examined by the scanning electron microscopy (SEM). The grain size was derived from SEM (Philips XL-30) inspection using an accelerating voltage of 10 kV and working distance of 10 mm. The average grain size, $d$, was measured by averaging direct grain measurements of over 100 grains per sample taken from SEM micrographs using image processing software (ImageJ). Since the produced SFO has high anisotropy, two length measurements were taken per grain to achieve a more accurate average. The two grain size measurements per grain resulted in an average radial and axial grain size. The crystallographic phase characterization was done by X-ray diffraction (XRD) using a PANalytical Empyrean

Diffractometer with Cu Kα X-ray source $\lambda_{K\alpha1}$=1.54056 Å $\lambda_{K\alpha2}$=1.54440 Å using 0.01313º step size.

**2.3 Magnetic measurements**

Magnetic hysteresis loops were measured using a Vibrating Sample Magnetometer (VSM) (Lakeshore 7400 Series) at room temperature. The external field values of up to 1.7 T were applied in order to attain the volume normalized magnetization, *M* [emu/cm$^3$] *vs.* applied field, H [Oe]. Magnetic induction B [G] was calculated using the relation $B = H + 4\pi M$. The sample was rotated in the magnetic field in order to attain M *vs.* H hysteresis loops at varying angle, θ. A Cube sample shape was used to negate geometrical influences on magnetic measurements and ensure varying θ measurements are comparable.

**2.4 Thermal measurements**

The CAPAD processed cylindrical samples with diameter of 19 mm and thickness of ~1.5 mm were cut into cubes of dimensions ~12 mm x 13 mm x 1.5 mm. The Laser Flash method was used for cross-plane thermal diffusivity and conductivity measurements, while the Hot Disk technique was used for in-plane thermal diffusivity and conductivity measurements.

*Laser Flash:* The cut samples were thinly coated with a graphite spray to help with transmittance of laser light and placed inside the LFA 447 instrument. The laser flash method is a transient method for measuring thermal diffusivity and operates by using a flash lamp as a heating source. The flash lamp heats one side of the sample while an InSb IR detector records the samples' backside temperature rise. The temperature rise is converted into an electrical signal and is used to calculate the thermal diffusivity, *α* of the sample. Thermal conductivity, *k* is calculated using *k = αρC$_p$*, where *C$_p$* is the specific heat of the sample and *ρ* is the density of the sample. The specific heat is measured separately using comparison with the sample with the known tabulated *C$_p$*. The cross-plane thermal conductivity was measured from 20 ºC to 100 ºC. Details of the Laser Flash measurement procedures have been reported by some of us elsewhere [13,14].

*Hot Disk:* The Hot Disk method is a transient method that allows one to simultaneously determine the thermal diffusivity and thermal conductivity [15-17]. The methods uses a sensor placed inside two sufficiently large pieces of a material under test. A constant pressure was applied during the measurements to minimize the influence of possible air gaps between the sensor and sample surfaces. The thermal characteristics are extracted from the temperature rise as a function of time recorded during the measurement. The physical principles and formulas for calculating the thermal diffusivity and thermal conductivity from the measured transient data can be found in Refs. [18-20].

3. Results and Discussion

**Figure 1b** shows an optical image of a typical CAPAD processed magnet next to a ruler. The samples are 19 mm in diameter and ~1.5 mm in thickness. A micrograph of the starting powder is shown in **Figure 1c**. The powder particles are relatively loose flakes with high aspect ratio. Average flake thickness (axial dimension) is 0.16 µm while average flake length (radial dimension) is 1.12 µm. Some particles are good examples of hexagonal prisms, which relates well to the hexagonal magneto-plumbite structure of $SrFe_{12}O_{19}$. **Figure 1d** is a micrograph of a fracture surface of a typical CAPAD processed magnet.

In order to examine the role of average grain size, $d$ and porosity, $P = 1 - \rho_{rel}$, we made two sets of samples. The first set has similar $\rho_{rel}$ (94.5 % to 96.4 %) with a range in $d$ (radial = 0.63 µm to 1.84 µm); the second had similar $d$ (radial = 0.63 to 0.78) but a range in $\rho_{rel}$ (86 % to 96 %). These two sets of samples are identified with shaded regions in **Figure 2a**, which plots average radial and axial grain sizes *vs.* $\rho_{rel}$ for the SFO samples.

In **Figure 2b** we present XRD data for the starting SFO powder, representative densified SFO samples (950°C, 104 MPa, 1 minute hold) and SFO ICSD standard (#202518). All XRD peaks in the powders, as well as densified samples, correspond to SFO, indicating no second phases in the starting powder and no detectable reaction in the CAPAD process. Comparison of the XRD peak intensities however does reveal preferential orientation of planes. The relative intensities of peaks of the randomly oriented powder match those of the ICSD standard while the XRD peak intensities of the densified SFO peaks differ in relative intensity. The peak intensities, which correspond to the (00X) planes such as (006) and (008) increase, while those that correspond to (XX0) peaks decrease in intensity. Materials with the hexagonal symmetry often have preferential growth directions so that the high aspect ratio flakes are formed. We attribute the preferential alignment in our samples to a combination of the high aspect ratio of the starting SFO powder flakes and preferential growth of grains during the CAPAD processing. The result is samples with a high aspect ratio grain morphology with larger grains in the in-plane direction compared to the cross-plane direction as mentioned before.

Magnetic hysteresis loop of a typical densified SFO (densified at 1000 °C) measured with H in the cross-plane direction is shown in **Figure 3a**. The M-H curve is relatively square as expected for a hard magnetic material. The energy product of the sample is $(BH)_{max} = 1.02$ MGOe which is in the range expected for a good ferrite magnet. The coercive field, $H_c$, is 2460 Oe. The magnetic saturation, $M_s$, which is calculated by subtracting the non-saturating slope (SFO is ferrimagnetic), is 380 emu/cm$^3$. The remnant magnetization, $M_r$, is 310 emu/cm$^3$. This yields an $M_r/M_s$ ratio of 0.82, which is much higher than 0.5 maximum for the isotropic magnets without grain alignment [21].

In order to further corroborate the XRD results that indicate preferential grain orientation, we conducted angle dependent magnetic measurements by varying the angle, θ between the magnetic field, H and samples' in-plane direction as shown in inset of **Figure. 3b. Figures 3b** and **c** show the B-H hysteresis curves (no demagnetization factor was used) as a function of angle, θ. A clear difference is observed in the B-H curves as the hysteresis loop is measured at different H orientations relative to the sample (θ = 0° (H in-plane), θ = 45° and θ = 90° (H cross-plane)). We used a cube of material allowing us to ensure no size/geometry effects or shape contributions to the magnetic measurement, which is confirmed by the samples having the same saturation magnetic induction. The measurements with H out-of-plane produces the largest remnant magnetic induction, $B_r$, followed by H 45° to plane and lastly by H in-plane at 3840 G, 3280 G and 2100, respectively. The $H_c$ from the B-H loop has a similar trend with H out-of-plane highest, followed by H 45° to plane and lastly by H in-plane at 1710 Oe, 1660 Oe and 1320 Oe, respectively. These measurements prove that the samples have preferentially oriented grains and that the easy axis of magnetization is in the cross-plane (θ = 90°) direction.

Electrical conductivity was measured for the samples processed at 800°C and 1000°C to determine whether phonons or electrons are the dominant contributors to heat conduction. The electrical conductivities for the samples processed at 800°C and 1000°C were $1.98 \times 10^{-7}$ $\Omega^{-1}$-cm$^{-1}$ and $3.1 \times 10^{-6}$ $\Omega^{-1}$-cm$^{-1}$, respectively. The electron contribution to the thermal conductivity was calculated using the Wiedemann-Franz law. It was determined to be 1.45 ×10$^{-10}$ W/(m·K) and 2.28 ×10$^{-9}$ W/(m·K), respectively. Such a low contribution signifies that phonons are the dominant heat carriers in $SrFe_{12}O_{19}$.

**Figure 4** shows the temperature dependence of thermal diffusivity (**Figures 4a** and **b**) and thermal conductivity (**Figures 4c** and **d**) in the in-plane and cross-plane directions for the set of samples with similar densities. The data correspond to the samples with various grain sizes. Note that the radial grain size is the characteristic grain size for the in-plane measurements while the axial grain size is used for the cross-plane. The in-plane measurements show decreasing thermal conductivity with temperature. For example, the decrease for the largest grain size sample (1.8 µm) is ~20% over this temperature range. This suggests that the thermal conductivity is limited to a large degree by the phonon Umklapp scattering in the in-plane direction. Umklapp scattering is an intrinsic phonon scattering mechanism originating from the crystal lattice inharmonicity. It is characterized by ~1/T dependence of the thermal conductivity on temperature [22,23]. By contrast, the cross-plane thermal conductivity does not change significantly (>2%) over this temperature range for all samples. Since the composition of the magnets does not change with measurement direction, these results suggest that grain boundary scattering is dominant scattering mechanism in the cross-plane direction. The boundary scattering becomes significant when the size of the structure becomes less than the intrinsic phonon mean free path [22,23]. In this case, the phonons do not travel far enough to encounter scattering from other phonons. Instead, they are scattered by the grain boundaries. The phonon transport regime dominated by the boundary scattering is characterized by a weak temperature dependence of the thermal conductivity. A larger number of grain boundaries in the cross-plane direction can also play a role in the measured thermal characteristics. The direction-dependent phonon thermal transport in our samples can be explained by the difference in the axial and radial grain sizes, with the latter being ~2 – 2.5 times large than the former.

The anisotropy in the grain sizes also causes a significant change in the magnitude of the *k* which ranges from 5.6 to 4.4 W/(m·K) for the in-plane and 2.65 to 2.25 W/(m·K) for the cross-plane direction at room temperature. The dependence of the thermal conductivity on the grain size for the samples with similar $\rho_{rel}$ can be seen in **Figure 5a**. The linear best fit lines are also plotted. The fit is good for the in-plane direction ($R^2 = 0.98$) and adequate for the cross-plane direction ($R^2$

= 0.84). The data show that the grain size dependence is stronger in the in-plane direction. The slope of the linear fit is nearly twice the cross-plane slope (0.94 for in-plane compared to the cross-plane slope, 0.55). A linear grain size dependence of $k$ (power law $k \propto d^1$) suggests the strong effect of the phonon – boundary scattering on heat conduction. The data of **Figure 5** together with **Figure 4** indicate that both Umklapp and grain boundary scattering are important in the in-plane direction while grain boundary scattering dominates in the cross-plane direction.

**Figure 5b** shows the porosity dependence of the set of samples with comparable grain sizes. As expected, the thermal conductivity decreases with porosity. The linear best fit lines are plotted and, again, the fit is better for in the in-plane direction ($R^2 = 0.99$) and adequate in the cross plane ($R^2 = 0.82$). For this set of samples, the porosity dependence is larger for the cross-plane direction. The slope of the line is -0.12 compared to -0.06 for in the in-plane direction. It is interesting to compare this porosity dependence with a classical theory for thermal conductivity in porous samples. It is known that the thermal conductivity of porous materials, $k_p$, can be written as [24, 25]:

$$k_p = k_s \frac{1+2P\frac{1-Q}{2Q+1}}{1-P\frac{1-Q}{2Q+1}}, \qquad \text{Eq. (1)}$$

where $P$ is the volume pore fraction, $k_s$ is the thermal conductivity of the non-porous solid and $Q$ is the ratio of $k_s$ to air ($Q = k_s/k_a$). In the case when $Q$ is relatively large, i.e. the thermal conductivity of the solid is large compared to air, Eq. (1) can be well approximated as linear over the considered porosity range. Using $k_s = 5.6$ W/(m·K) for SFO and 0.0257 W/(m·K) for air, the slope in our porosity range given by Eq. (1) is -0.07 which is very close to -0.06 we measured in the in-plane direction. In the cross-plane direction the slope (-0.12) is almost twice the predicted slope. This means that the classical theory, expressed by Eq. (1), captures the porosity dependence accurately when the characteristic grain size is relatively large (as in the in-plane direction in our samples). However, this theory underestimates the effect of porosity when the characteristic grain size is in the nanometer to sub-micrometer range. The latter can be attributed to the strong phonon boundary scattering effects in the samples with smaller grain sizes. Similar discrepancy between the experiments and classical theory of the effects of porosity has been observed before in nanocrystalline samples synthesized by the same CAPAD technique [9].

The obtained results have important implications for practical application of permanent magnets. As mentioned previously, PMs are being increasingly used in environments where they are exposed to elevated temperatures, which decrease the magnetization. The anisotropy of a magnet can be used to optimize both magnetic performance and provide an effective direction for heat removal. The magnet will provide the highest magnetic field along the easy axis (as demonstrated in **Fig. 3**), which in the example here is in the cross-plane direction. In order to maintain the temperature as low as possible, and maintain the best magnetic performance, one should cool the PM in the in-plane direction (along directions of largest grains), since the thermal conductivity is about twice the cross plane direction (as demonstrated in **Fig. 4**).

## 4. Conclusions

In summary, we produced dense SFO magnets that have high aspect ratio grain morphology resulting in preferential alignment of crystallographic planes. The magnets have the magnetic easy axis aligned perpendicular to their largest surface area (cross-plane) to maximize their magnetic performance. They have higher thermal conductivity in the in-plane direction providing opportunity for effective cooling. Temperature and direction dependent thermal conductivity indicate that both Umklapp and grain boundary scattering are important in the in-plane direction (with relatively large characteristic grain size) while grain boundary scattering dominates in the cross-plane direction (with relatively small characteristic grain size). These results should translate well to other nanostructured permanent magnets and thus have important implications of the thermal management of permanent magnets in demanding applications such as generators and motors.

*Acknowledgements:* The work was supported as part of the SHINES, an Energy Frontier Research Center funded by the US Department of Energy, Office of Science, Basic Energy Sciences under Award No. SC0012670. The authors thank Edberg Uy for help with experiments.

**FIGURES**

**Figure 1:** a) A schematic of the high aspect ratio microstructure of permanent magnets. The magnetic easy axis is aligned cross-plane, maximizing surface area in the highest magnetization direction. Larger grain sizes in the in-plane direction enhance the thermal conductivity. b) An optical image of CAPAD processed SFO magnet. The sample dimensions are 19 mm in diameter and ~1.5 mm in thickness. c) A micrograph of the starting SFO powder. The powder particles are relatively loose flakes with the high aspect ratio. d) A micrograph of a fracture surface of a typical CAPAD processed SFO magnet.

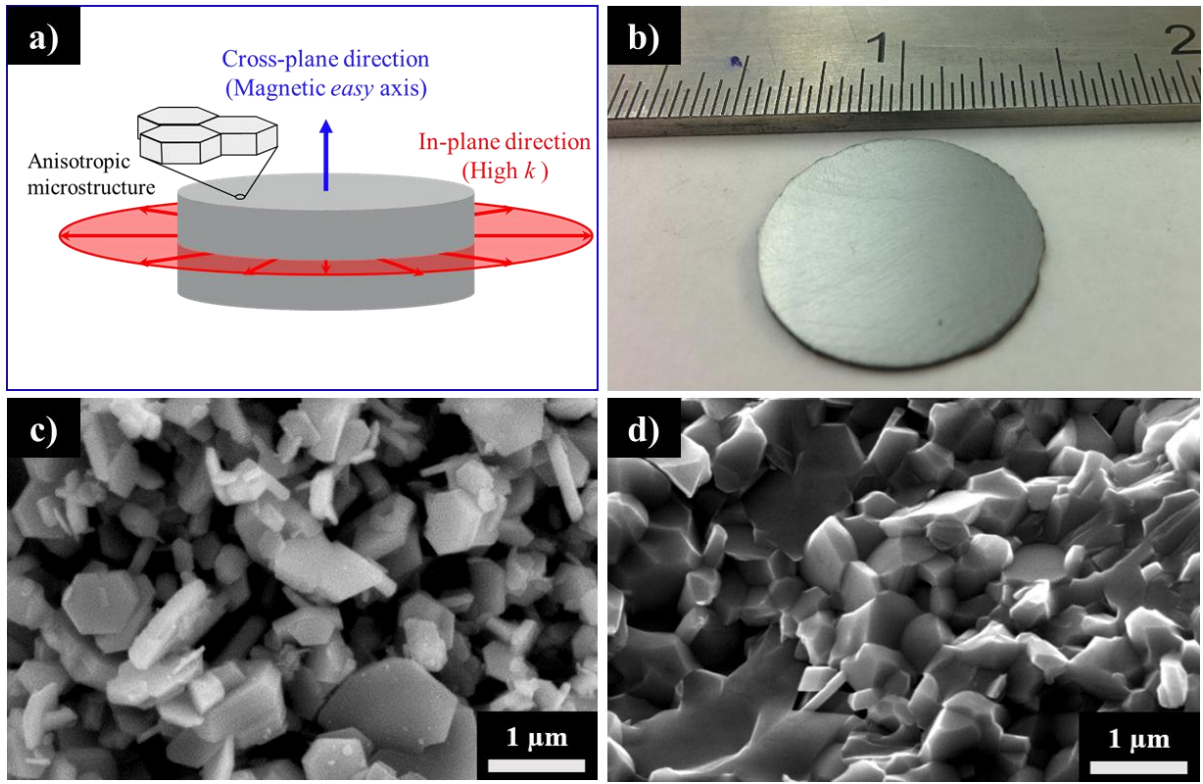

**Figure 2:** Microstructural characteristics of CAPAD processed SFO samples. a) The average radial and axial grain sizes *vs.* relative density. b) XRD profiles for the starting SFO powder and a representative densified SFO sample (950°C, 104 MPa, 1 minute hold). The profile of ICSD standard (#202518) is plotted for comparison. Comparison of the XRD profiles reveals preferential orientation of the planes in the bulk sample.

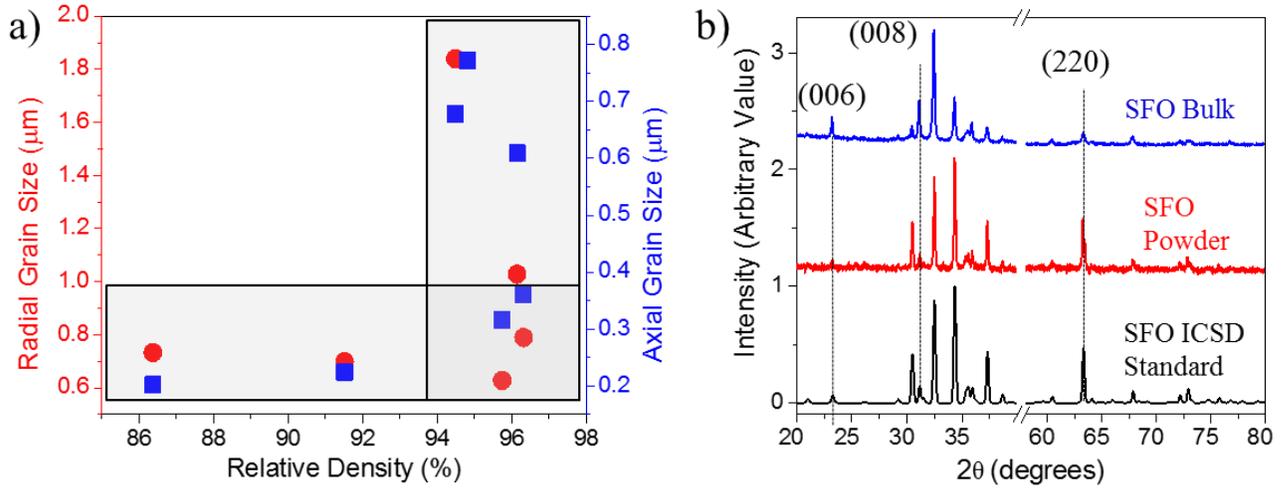

**Figure 3:** Magnetic properties of CAPAD processed SFO magnets. a) Magnetic hysteresis loop of a typical densified SFO (densified at 1000°C) measured with H in the cross-plane direction. b) B-H hysteresis curves as a function of the measurement angle, θ. The inset shows the relation between θ and the surfaces of the magnets. c) A magnification of the B-H measurements in the second quadrant, clearly showing orientation effects.

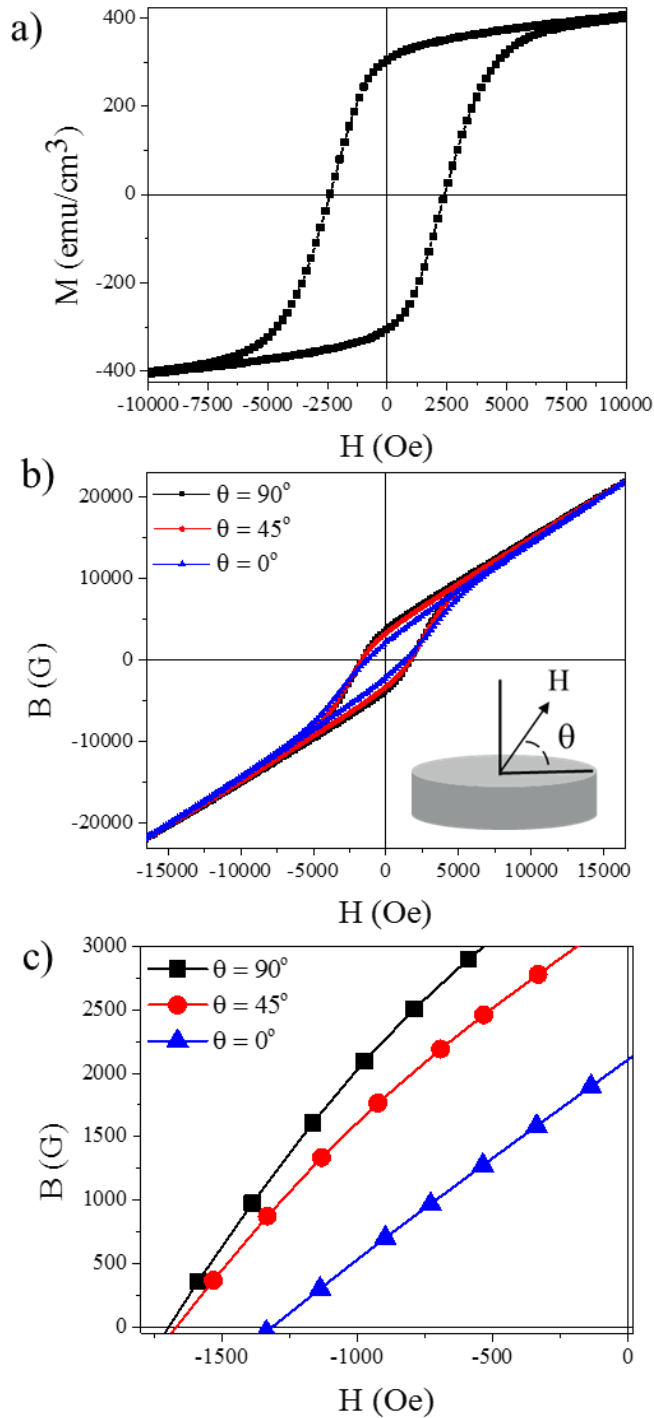

**Figure 4:** Thermal properties of CAPAD processed SFO magnets with similar relative densities: a) thermal diffusivity in the in-plane direction; b) thermal diffusivity in the cross-plane direction; c) thermal conductivity in the in-plane direction; d) thermal conductivity in the cross-plane direction.

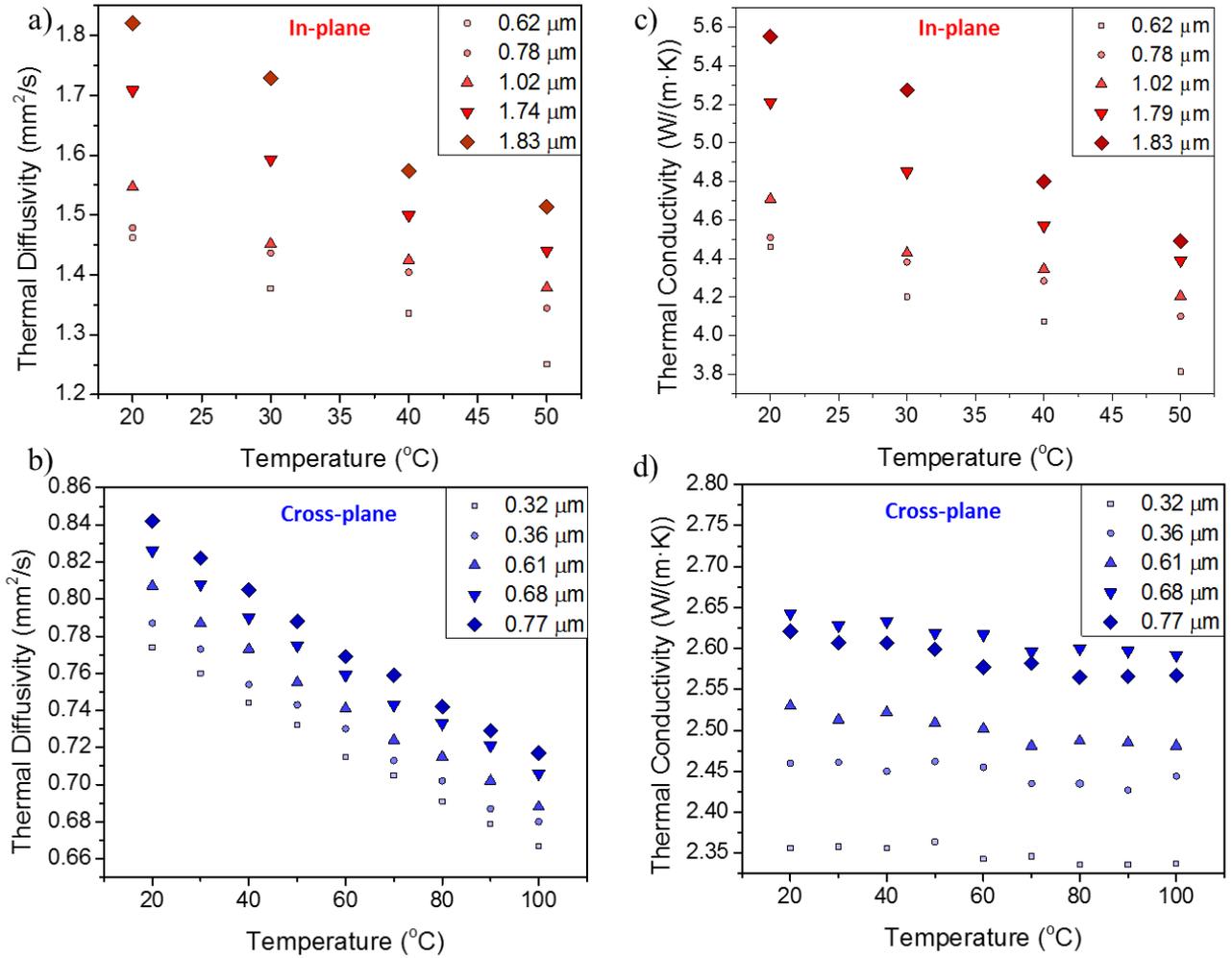

**Figure 5:** a) The dependence of thermal conductivity on grain size for samples with similar relative density in the in-plane and cross-plane direction. b) The dependence of thermal conductivity on porosity for samples with the similar grain size in the in-plane and cross-plane direction.

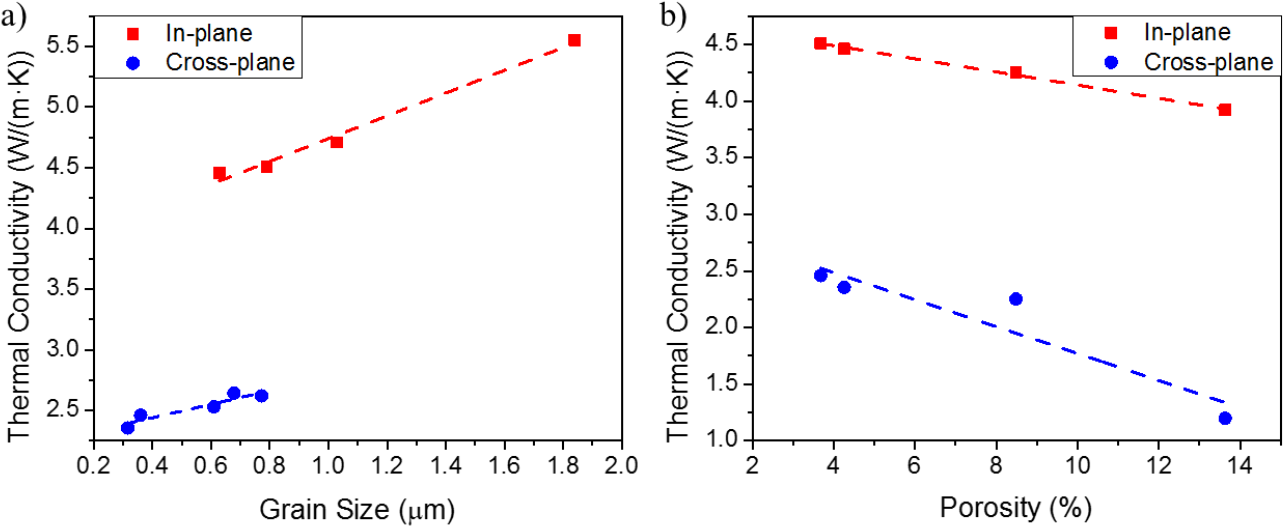